\documentclass{PoS}

\newcommand{\Ord}{\mathcal{O}}
\newcommand{\csw}{c_{SW}}
\newcommand{\ev}[1]{\left\langle #1 \right\rangle}
\newcommand{\fdet}[1]{\textnormal{det}\left(D_{SW}(#1)[U]\right)}
\newcommand{\Tr}[1]{\textnormal{Tr}\left(#1\right)}
\newcommand{\dens}{\mathcal{D}}

\newcommand{\refc}[1]{(\ref{#1})}

\title{Towards the $N_f=2$ deconfinement transition temperature with $\Ord(a)$ improved Wilson fermions}

\ShortTitle{The $N_f=2$ deconfinement transition with $\Ord(a)$ improved Wilson fermions}

\author{\speaker{Bastian B. Brandt} and Hartmut Wittig\\
        Institut f\"ur Kernphysik, Johannes Gutenberg-Universit\"at Mainz, \\
        Johann Joachim Becher-Weg 45, 55099 Mainz, Germany\\
        E-mail: \email{brandt@kph.uni-mainz.de}, \email{wittig@kph.uni-mainz.de}}
\author{Owe Philipsen and Lars Zeidlewicz\\
        Institut f\"ur Theoretische Physik, Goethe-Universit\"at, \\
        Max-von-Laue-Str. 1, 60438 Frankfurt am Main, Germany\\
        E-mail: \email{philipsen@th.physik.uni-frankfurt.de}, \email{zeidlewicz@th.physik.uni-frankfurt.de}}

\abstract{A lot of effort in lattice simulations over the last years has been
	devoted to studies of the QCD deconfinement transition. Most
	state-of-the-art simulations use rooted staggered fermions,
	while Wilson fermions are affected by large systematic uncertainties, such as coarse lattices
	or heavy sea quarks. Here we report on an
	ongoing study of the transition, using two degenerate flavours of
	nonperturbatively $\Ord(a)$ improved Wilson fermions. We start with $N_{t}=12$ and 16
	lattices and pion masses of 600 to 450 MeV,
	aiming at chiral and continuum limits with light quarks.}

\FullConference{The XXVIII International Symposium on Lattice Field Theory, Lattice2010\\
		June 14-19, 2010\\
		Villasimius, Italy}

\begin{document}

\section{Introduction}

The high temperature transition from hadronic matter, surrounding us in the universe, to the deconfined state of the quark gluon plasma plays an important role in the current state of research in nuclear and particle physics, from the theoretical as well as from the experimental point of view. Since a lot of experimental effort is being invested in heavy ion collision programmes (CERN SPS, LHC and RHIC), investigating the properties of the phase transition and the deconfined phase by theoretical means becomes mandatory. This particularly concerns the order of the phase transition and the role of the related chiral symmetry restoration. Since the quark gluon plasma is still a strongly coupled system for intermediate temperatures,  a perturbative treatment of the plasma in that region is not valid and lattice QCD is the preferred tool to study the transition.

Most of the simulations so far were performed using staggered
fermions, having the advantage of being numerically cheap compared to
other fermion discretisations. Recent results can be found in
refs.\,\cite{hotQCD,BW-group}. However, in order to realise the
physical number of quark flavours, fractional powers of the staggered
quark determinant have to be taken (the ``rooting'' procedure). The
long-standing controversy about the validity of rooting (see
e.g. \cite{rooting}) is on-going. Furthermore, the definition of the
physical pion mass in the rooted staggered approach is ambiguous: At
this conference, it was debated whether it is correct to refer to the
lightest pion mass determined in simulations as the physical pion
mass, or whether one should use the averaged pion mass as an estimate
\cite{Kanaya}. This demonstrates that a cross-check of the
staggered results is needed using other fermionic
discretisations. Several groups have already started to perform
simulations with two flavours of non-perturbatively $\Ord(a)$ improved
Wilson fermions, of the Sheikholeslami-Wohlert type
\cite{QCDSF,WhotQCD} as well as with the maximally twisted mass type
\cite{tmft}. All these simulations still suffer from uncontrolled
systematic effects, namely unphysically large pion masses and
lacking continuum extrapolations, even though recent results significantly improved
the situation \cite{QCDSF_new,tmft_new}.

In this paper we report on an ongoing study of the $N_f=2$ phase transition with non-pertur\-batively $\Ord(a)$-improved Wilson fermions, starting at $N_t=12$ and pion masses $m_\pi \lesssim 600$ MeV. The aim is to extract the transition temperature and the order of the transition in the chiral limit, which is still not settled until today. There are two possible scenarios: In the first scenario, the chiral critical line in the $\{m_{u,d},m_{s},T\}$-parameter space never reaches the $m_{u,d}=0$ axis, while the second implies the existence of a tricritical point at $m_{u,d}=0$. This tricritical point has to extend into the direction of finite chemical potential as a critical line.
Thus our $N_f=2$ simulation addresses a question on the phase diagram of the $N_f=2+1$ theory and is important for the phase diagram at finite density as well. First results suggest that the second scenario is favored by lattice data \cite{QCDSF,QCDSF_new}, but systematics have to be improved in order to make a precise statements.

\section{Setup of the simulations}

\subsection{Lattice action and update algorithm}

In our simulations, we work with two degenerate flavours of nonperturbatively $\Ord(a)$ improved Wilson fermions, using the Sheikholeslami-Wohlert lattice Dirac operator \cite{SW-action}
\begin{equation}
 \label{eq-sw-op}
D_{SW} = D_W + \csw \: \frac{i\:a\:\kappa}{4} \: \sigma_{\mu\nu} \: \hat{F}^{\mu\nu} \equiv D_W + \csw \: M_{SW} \; .
\end{equation}
Here $D_W$ is the usual Wilson Dirac operator, $\kappa$ is the hopping parameter, $\sigma_{\mu\nu}$ the totally antisymmetric tensor and $\hat{F}^{\mu\nu}$ the 'clover leaf' representation of the gluonic field strength tensor on the lattice. To ensure $\Ord(a)$ improvement for spectral quantities, the clover coefficient $\csw$ has to be tuned properly with $\beta$, using the well known interpolation formula for $5.2<\beta<12.0$ from \cite{npcsw}.

For simulations with large lattices and small pion masses, an efficient algorithm is needed. Here we use the deflation accelerated DDHMC algorithm, introduced by L\"uscher \cite{DDHMC}. This algorithm is nowadays widely used in large-scale lattice simulations and known to work well for large lattices up to $128\times 64^3$ and pion masses down to 250 MeV (see e.g. \cite{CLS,CLS_wiki}). Although the algorithm performs well for a variety of lattice sizes it introduces some limitations for simulations at non-zero temperature. In particular, we are restricted to lattices sizes $N_{t,s}=8,12,16,20,24,\ldots$ in each direction with the current version of the algorithm, due to the division of the lattice into an even number of DDHMC domains. Following these geometrical constraints and the fact that $\csw$ is known nonperturbatively only for $\beta>5.20$, we see that we have to go to lattices with $N_t\geq12$, in order to simulate at small pion masses.

\subsection{Observables and strategy}

We investigate the behaviour of three observables around the critical temperature, the average plaquette $\ev{P}$, the real part of the Polyakov loop $\textnormal{Re}\left[\ev{L}\right]$ and the chiral condensate $\ev{\bar{\psi}\psi}$. In \cite{hw-89} it was found that observables constructed from smeared gauge links can lead to a more pronounced signal in investigations of the phase structure. Motivated by this reference we also computed the Polyakov loop using 5 levels of APE smearing with a weight of 0.5 for the staples. We define the generalised susceptibilities $\chi(O)$ by
\begin{equation}
 \label{susz}
\chi(O) \equiv N_s^3 \: \left( \ev{O^2} - \ev{O}^2 \right) ,
\end{equation}
where $O$ is any of the observables above and $N_s$ the spatial lattice size. These generalised susceptibilities should show a notable peak at the transition point. In addition, the behaviour of the peak under a change in the spatial volume is governed by the corresponding critical exponents, giving information about the order of the transition.

The temperature $T$ of the system is related to the temporal extent via $T=1/(N_t\:a)$, where $a$ is the lattice spacing. There are in general two possible procedures to perform a temperature scan. The first is to fix the lattice spacing $a$ and vary the temporal extent $N_t$ of the lattice. This procedure has the advantage of leaving the scale invariant, but the clear disadvantage is that the resolution is limited. This is particularly crucial around the critical temperature and even made worse by the use of DDHMC, because of the restrictions on $N_t$ discussed above. The other method which we use here is to vary the lattice spacing $a$, while keeping the temporal extent $N_t$ fixed. One scheme to locate the transition is used by the QCDSF-DIK collaboration: they cross the transition point by varying the hopping parameter $\kappa$, which is related to the bare quark mass, while keeping the lattice coupling $\beta$ fixed \cite{QCDSF}. The scheme we use here is to fix $\kappa$ while varying $\beta$, enabling us to get a fine resolution around $T_c$. Furthermore we are able to employ the multi-histogram method discussed in the next subsection.

\subsection{Multi-Histogram methods with an explicitly $\beta$ dependent fermion determinant}
\label{ch-mhist}

Multi-Histogram methods are a tool known for quite some time that allows to use a set of measurements at some values $\{\beta_a\}$ for an interpolation of the expectation values of observables at some different value $\beta$ \cite{mh-ref}. The method is applicable to theories including fermions, as long as the Dirac operator does not depend explicitly on the coupling $\beta$. For unimproved Wilson fermions this is true while for $\Ord(a)$ improved Wilson fermions the determinant becomes $\beta$-dependent. For the Sheikholeslami-Wohlert Dirac operator \refc{eq-sw-op}, the $\beta$-dependence is encoded in the tuning of $\csw$, spoiling the applicability of the usual method. In a future publication \cite{mhist-paper} we will show in detail how the multi-histogram method can be extended to the case of $\Ord(a)$ improved Wilson fermions. Here we only present an abridged discussion of the main ideas.

For a $\beta$-dependent Dirac operator and two degenerate flavours, the weight of a given configuration $[U]$, generated at $\beta_0\in\{\beta_a\}$ for a measurement at $\beta$ is given by
\begin{equation}
 \label{eq-dens}
P(\beta)[U] = e^{-\left(S_g(\beta)[U]-S_g(\beta_0)[U]\right)} \: R(\beta,\beta_0)[U] \: P(\beta_0)[U] \: ,
\end{equation}
where $P(\beta_0)[U]$ is the statistical weight of the configuration in the $\beta_0$-ensemble and we have introduced an abbreviation for the ratios of fermion determinants at different $\beta$'s:
\begin{equation}
 \label{det-rat}
R(\beta,\beta_0)[U] \equiv \left(\frac{\fdet{\beta}}{\fdet{\beta_0}}\right)^2 \;.
\end{equation}
To have an efficient method, it is necessary to evaluate these ratios without the need for full knowledge of the determinant, which would demand a new calculation for each value of $\beta$. Using eq.~\refc{eq-sw-op}, we can rewrite $D_{SW}(\beta)$ as
\begin{equation}
D_{SW}(\beta)[U] = D_{SW}(\beta_0)[U] + \Delta\csw(\beta,\beta_0) \: M_{SW}[U] ,
\end{equation}
with $\Delta\csw(\beta,\beta_0) \equiv \csw(\beta)-\csw(\beta_0)$. Inserting this expression into eq.~\refc{det-rat} leads to
\begin{equation}
 \label{det-rat2}
R(\beta,\beta_0)[U] = \left(\textnormal{det}\left( 1 + \Delta\csw(\beta,\beta_0) \: D_{SW}^{-1}(\beta_0)[U] \: M_{SW}[U] \right) \right)^2 .
\end{equation}
In typical $\beta$-scans $\Delta\csw\lesssim\Ord(10^{-2})$ over the whole range of $\beta$, which suggests that it may be sufficient to expand eq.~\refc{det-rat2} in $\Delta\csw$ to approximate the ratio of the two determinants. A short calculation yields
\begin{equation}
 \label{det-rat3}
\begin{array}{rl}
\displaystyle R(\beta,\beta_0)[U] & \displaystyle = 1 + 2\:\Delta\csw(\beta,\beta_0) \: \Tr{D_{SW}^{-1}(\beta_0)\:M_{SW}[U]} + \Ord(\Delta\csw^2) \vspace*{2mm} \\
 & \displaystyle \equiv 1 + 2\:\Delta\csw(\beta,\beta_0) \: R_{lin}(\beta_0)[U] + \Ord(\Delta\csw^2) .
\end{array}
\end{equation}
To apply the multi-histogram method to the full set $\{\beta_a\}$, one has to introduce a global density of states $\dens$ as in \cite{mh-ref}. This is not possible without full knowledge about the determinant at each $\beta_a$. A possibility to circumvent this problem, is to introduce a reference point $\beta_R$ for the fermionic part, together with the corresponding global density $\dens(\beta_R)$. In this way, the multi-histogram method can be employed using the linearised expression for $R(\beta,\beta_0)$ discussed above, provided that the coefficients multiplying the higher-order terms are not larger than one. It is then sufficient to compute the quantity $R_{lin}(\beta)$ for each $\beta_a$-value of the scan. The combination $D_{SW}^{-1}M_{SW}$ which enters $R_{lin}$ can be determined with relatively little computational overhead using stochastic estimators. Furthermore, $R_{lin}$, being the derivative of $R(\beta,\beta_0)$ at coupling $\beta_0$, provides some control over the quality of the linearised ansatz for the ratio of determinants.

\section{Simulation results}
\label{results}

\begin{table}[t]
\centering
\begin{tabular}{c|cccccccc}
\hline
\hline
scan & Lattice & Block size & $\kappa$ & $\beta$-range & $\tau$ & $\tau_{int}[P]$ & $f_{meas}$ & Statistics \\
\hline
$A$ & $12\times24^3$ & $6^4$ & 0.13595 & $5.27-5.32$ & 2.0 & $\Ord(30)$ & 1 & $\Ord(25000)$ \\
$B$ & $16\times32^3$ & $8^4$ & 0.13650 & $5.40-5.55$ & 2.0 & $\Ord(10)$ & 2 & $\Ord(4000)$ \\
\hline
\hline
\end{tabular}
\caption{Run parameters for scans $A$ and $B$. We show the DDHMC block size, the Monte Carlo time $\tau$ of the trajectories, the measurement frequency $f_{meas}$ and the integrated autocorrelation time $\tau_{int}$ of the plaquette $P$.}
\label{tab1}
\end{table}

So far our simulations where done on two different lattices, namely $12\times24^3$ (which we refer to as scan $A$) and $16\times32^3$ (scan $B$). The corresponding parameters are given in table \ref{tab1}.

\begin{figure}[t]
\centering
\begin{minipage}{.45\textwidth}
\centering
\includegraphics[angle=-90, width=.9\textwidth]{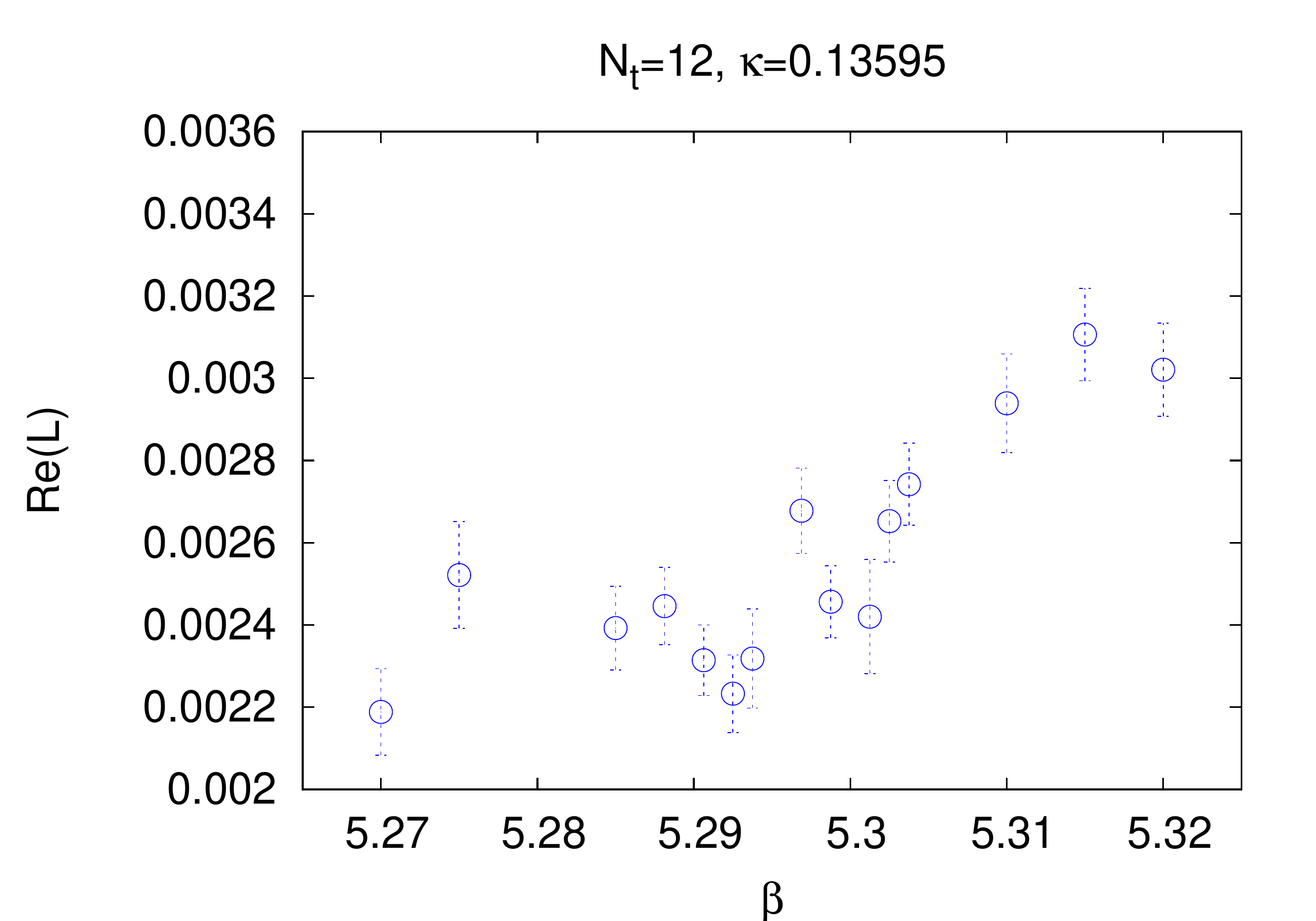}
\newline
\includegraphics[angle=-90, width=.9\textwidth]{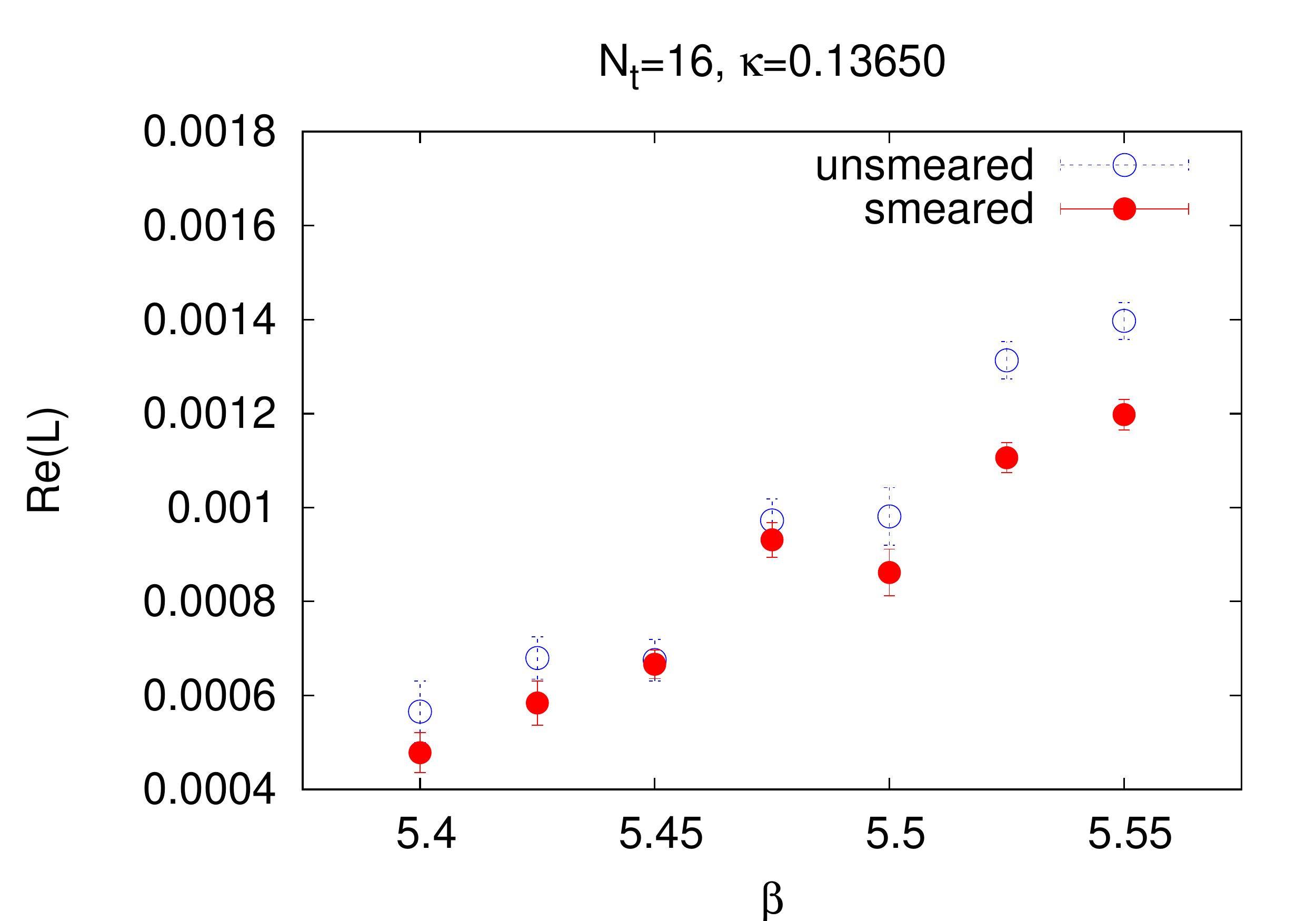}
\end{minipage}
\begin{minipage}{.45\textwidth}
\centering
\includegraphics[angle=-90, width=.9\textwidth]{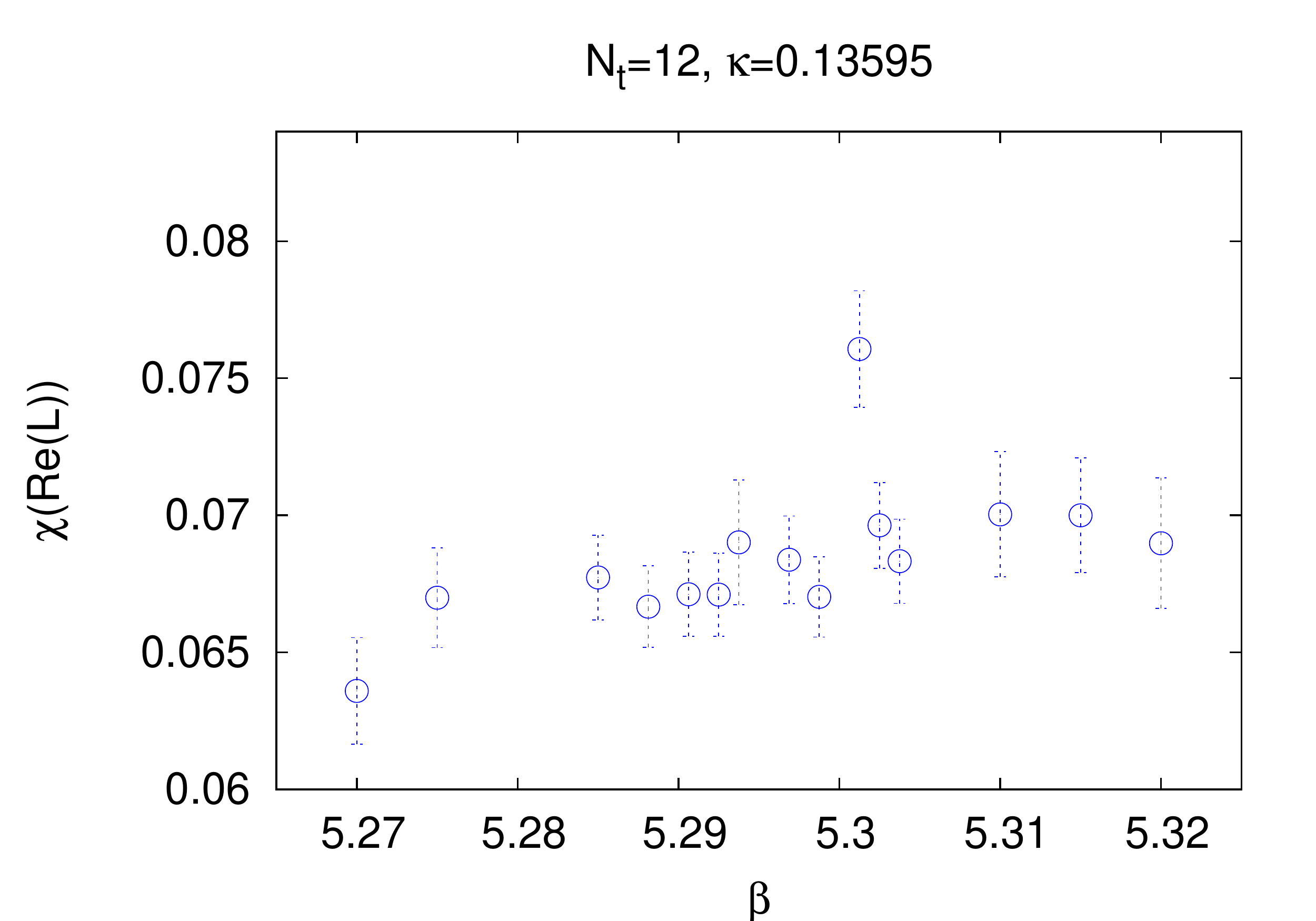}
\newline
\includegraphics[angle=-90, width=.9\textwidth]{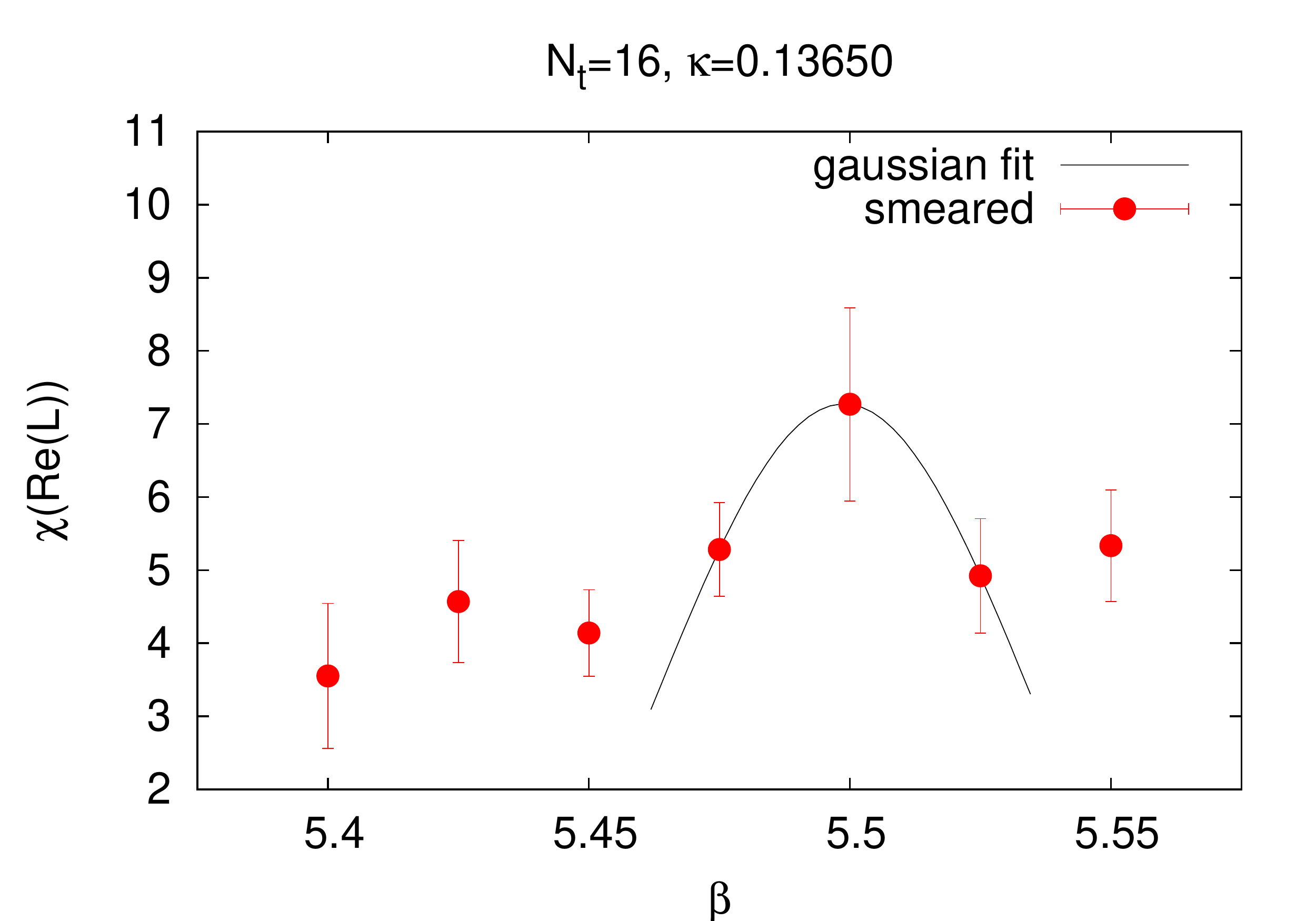}
\end{minipage}
\caption{{\bf TOP:} Plot of the results for scan $A$. {\bf {\it Left:}} Real part of the Polyakov loop. {\bf {\it Right:}} Corresponding susceptibility. {\bf BOTTOM:} Plot of the results for scan $B$. {\bf {\it Left:}} Real part of the Polyakov loop in the unsmeared and APE-smeared version. The smeared results are rescaled such that we could show them in a single plot. {\bf {\it Right:}} The susceptibility of the APE-smeared Polyakov loop, together with a gaussian fit at the peak position.}
\label{fig1}
\end{figure}

Scan $A$ was performed by setting the hopping parameter $\kappa$ to the value where the transition in the real part of the Polyakov loop was observed for $\beta=5.29$ in \cite{QCDSF}. On the top of figure \ref{fig1} we show the behaviour of $\textnormal{Re}\left[\ev{L}\right]$ for scan $A$. The data are consistent with a phase transition at $\beta_c=5.301(3)$, where we see a strong increase in the signal and a peak in the susceptibility. By direct comparison, we see that the resulting transition temperature is slightly higher than the one obtained in \cite{QCDSF}. A similar observation was made when comparing the physical transition temperatures from \cite{QCDSF} to a simulation using twisted mass fermions \cite{tmft_new}. Since for the relatively large pion mass of roughly 600 MeV the transition is hardly a sharp phase transition but a broad crossover, it is to be expected that estimates for the transition temperature derived from different observables or directions in parameter space differ. Since we did not perform any scale setting runs at $T=0$, we do not give an estimate for the transition temperature in physical units.

Conceptually scan $A$ only served as a feasibility test for our algorithm and our measurement routines and an easy tuning case for the simulations at $N_t\geq16$. In addition it turned out that this scan is at the lower boundary of the working region of the algorithm in the sense that the small lattice size limits the possibilities to tune the DDHMC parameters. One of the consequences are the relatively large integrated autocorrelation times for the plaquette as can be seen in table \ref{tab1}, which are also reproduced by other observables. These large autocorrelation times are most likely due to the relatively small DDHMC blocks of $6^4$, limiting the number of active links during the HMC evolution (see \cite{DDHMC} for further explanations). Furthermore the pion mass of scan $A$ is still very large. We aim at much smaller masses where we also expect a stronger signal for the transition.

Scan $B$ is our first scan at lighter pion mass and our first attempt for simulations at $N_t=16$. We show the behaviour of $\textnormal{Re}\left[\ev{L}\right]$ together with the smeared version $\textnormal{Re}\left[\ev{L}_{sm}\right]$, and the susceptibility of the latter, together with a Gaussian fit to 3 points around the peak position, at the bottom of figure \ref{fig1}. The peak position is obtained from the fit as $\beta_c=5.499(2)$. Fortunately there already exists a run for $T=0$ with the parameters $\beta=5.50$ and $\kappa=0.13650$ \cite{CLS}, leading to a transition temperature in physical units of roughly $T_c(m_{\pi}=\:457\:\textnormal{MeV},a=0.060\:\textnormal{fm})\approx206$ MeV.
Since the determination of the scale is still in progress, we do not give an estimate for the error, but it should be kept in mind that the systematical error might still be large. It is important to note that the peak is reproduced by all other observables as well.

\section{Conclusions and outlook}

In this proceedings article, we gave a first overview about our ongoing study of the QCD deconfinement transition. We introduced the setup and the main observables to extract the transition temperature and an extension of the multi-histogram method to increase the resolution around $T_c$. We discussed first results, including the test scan at $N_t=12$ and a first scan at $N_t=16$.
The transition temperature for the latter is at around 206 MeV, but at a still quite large pion mass of roughly 450 MeV.
The peak in the susceptibilities is reproduced by all observables. The $N_t=16$ scan is still extended at the moment to improve the signal and a final analysis is postponed to the next publication. We are going to enlarge our set of scans at $N_t=16$ and also aim at even larger values of $N_t$. After mapping the critical trajectories $\beta_c(\kappa,N_t)$, we are going to do simulations on different volumes in order to perform a finite size scaling analysis. In the long run, we aim to do an extrapolation to the chiral limit and a separated study of the continuum limit supported by several temporal lattice extents and several pion masses for each $N_t$.

\acknowledgments

The simulations where done on the WILSON cluster at the University of Mainz (see \cite{CLS}) and on JUGENE at FZ Juelich under NIC Grant Nr. 3330. We are indebted to the institutes for these facilities. We also like to thank H.B. Meyer for many fruitful discussions. B.B. is funded by the DFG via SFB 443. L.Z. is supported by DFG PH 158/3-1.

\end{document}